\documentclass[%
reprint,
showpacs,
floatfix,
amsmath,amssymb,
aps,
prb,
floatfix,
]{revtex4-1}

\usepackage{color}
\usepackage{graphicx}
\usepackage{dcolumn}
\usepackage{bm}

\newcommand*{\citen}[1]{%
  \begingroup
    \romannumeral-`\x 
    \setcitestyle{numbers}%
    \cite{#1}%
  \endgroup   
}

\DeclareMathOperator{\sgn}{sgn}

\def\eeq{\relax}
\def\beq#1#2\eeq{\begin{equation}\label{#1}#2\end{equation}}
\def\bal#1#2\eal{\begin{align}\label{#1}#2\end{align}}
\def\bse#1#2\ese{\begin{subequations}\label{#1}#2\end{subequations}}

\begin{document}


\title{Retrieval method for the bianisotropic polarizability tensor of Willis acoustic scatterers} 

\author{Xiaoshi Su}
 \email{xiaoshi.su@rutgers.edu}
 \affiliation{Department of Mechanical and Aerospace Engineering, Rutgers University, Piscataway, NJ 08854}
\author{Andrew N. Norris}%
 \email{norris@rutgers.edu}
\affiliation{Department of Mechanical and Aerospace Engineering, Rutgers University, Piscataway, NJ 08854}

\date{\today}

\begin{abstract}
Acoustic materials displaying coupling between pressure and momentum  are known as Willis materials. The simplest Willis materials are comprised of   sub-wavelength scatterers that couple monopoles to  dipoles and {\it vice versa}, with the interaction  defined by a  polarizability tensor. We propose a  method for retrieving the polarizability tensor for sub-wavelength Willis acoustic scatterers using a finite set of scattering amplitudes. We relate the polarizability tensor to standard T-matrix and S-matrix scattering formalisms.  This leads to an explicit method for retrieving the components of the polarizability tensor in terms of a small set of  scattered pressure data in the near- or  far-field.  
Numerical examples  demonstrate the retrieval method for one and two dimensional configurations. 
\end{abstract}

\maketitle


\section{\label{intro}Introduction}
Acoustic metamaterials comprised of sub-wavelength inclusions have attracted much attention due to their extraordinary acoustical effects. By proper design of the inclusions one can achieve a variety of exotic effective material properties such as negative bulk modulus and density,~\cite{Liu000,Li04a,Fang06} zero index,~\cite{Huang11,Dubois2017a} anisotropic and pentamode properties.~\cite{Popa2016,Su2017a,Hedayati2017} Numerous applications have been demonstrated, e.g. acoustic cloaking,~\cite{Fang11,Chen2017} negative refraction,~\cite{Zhang2004,Hladky-Hennion13} and nonreciprocal transmission.~\cite{Liang2010,Fleury2014b} The medium of particular interest in this paper is known as a Willis material, which requires strain-velocity cross-coupling in elasticity or pressure-velocity coupling in acoustics.~\cite{Willis1980,Willis1981,Milton07,Norris12,Muhlestein2016} 

It is well-known that monopole and dipole moments are dominant sources of the scattered field for sub-wavelength scatterers. For Willis type acoustic scatterers, the scattering requires cross-coupling between pressure and velocity fields.\cite{Sieck2017} Recently, \citet{Quan2018} described this type of coupling in the context of acoustic bianisotropic polarization due to its similarity to  bianisotropy in electromagnetics. One interesting finding in their work is that both the pressure and velocity fields can produce monopole and dipole responses. In some circumstances, the cross-coupling induced scattering can be dominant for properly designed sub-wavelength scatterers. The interaction between pressure and velocity fields makes the scattering problem more complicated, while also presents novel wavefront manipulation possibilities that conventional metasurfaces do not have.~\cite{Zhao2013,Xie2014,Li2015,Dubois2017,Liu2017,Su2018} For example, \citet{Koo2016} took advantage of the pressure-velocity coupling to achieve simultaneous control of transmitted and reflected wavefronts. \citet{Li2018} proposed a systematic design approach to improve the efficiency of acoustic metasurfaces using bianisotropic unit cells. The aforementioned two examples are based on the  cross-coupling in uniaxial materials, the coupling in two- and three-dimensional (2D and 3D) Willis scatterers contain richer physics and can be potentially used in unprecedented applications. In order to design such scatterers, a rigorous and handy retrieval method for the polarizability tensor needs to be established. There are several retrieval methods available in the literature of electromagnetics.~\cite{Arango2013,Mirmoosa2014,Asadchy2014,Liu2016} However, the retrieval method must be redeveloped in acoustics due to the difference in the constitutive equations.

Examples of acoustic retrieval of Willis properties are limited.  Thus, \citet{Muhlestein2017a} established a method for extracting the effective bulk modulus and mass density for a one-dimensional (1D) Willis material.
More recently, \citet{Quan2018} provided a robust method for retrieving the polarizability tensor for 2D structures by directly calculating the monopole and dipole moments based on the orthogonality of each mode.  The latter method requires knowledge of the scattered field in all directions for different types of wave incidence, i.e.\ an infinite data set.  The polarizability tensor contains a finite number of independent elements, 3 in 1D and 6 in 2D, and it therefore should be possible to retrieve these using   finite data sets of similar size.  The purpose of this paper is to present a simple and efficient  approach for obtaining the Willis polarizability tensor in both 2D and 1D that requires only limited data. We do this by first interpreting the   polarizability tensor in terms of a scattering  (or T-) matrix. 
This places the Willis polarizability  in the context of standard scattering theory in which the T-matrix has full coupling.  The scattering approach also implies bounds on the polarizability in a natural way by using the relation between  the T-matrix and the  S-matrix. 
The bounds are a consequence of energy conservation which implies that the eigenvalues of the S-matrix must be of magnitude less than or equal to unity.

Explicit formulae are given for retrieving the polarizability tensor from finite sets of scattering data and are illustrated using numerical simulations. The method developed here is significantly  simpler than that of \citet{Quan2018}  in that it only requires a few probes of the scattered pressure in the near- or far-field for several plane wave excitations. For a Willis scatterer in 2D free field, there are nine polarizability components in total but only six  are independent. One only needs to simulate plane wave excitations along the $\pm x$- and $\pm y$-directions, and probe the far-field scattered pressure   along the $\pm x$- and $\pm y$-directions. Note that the choice of the incident directions and probe locations are not unique. For a Willis scatterer in an 1D waveguide, the total number of  polarizability components reduces to four with three independent ones. In this case one only needs   incidence along opposite directions and data for  the scattered pressure on both sides. Such an easy-to-implement method can drastically facilitate the optimization process during the inclusion design.  It also offers the experimentalist a simple method to measure the polarizability.  Two examples will be presented to demonstrate our retrieval method. It should be noted that higher order multipole moments also exist in the scattered field, but their contributions are negligible for deep sub-wavelength scatterers. We will see from the two examples that the curves obtained using the method developed here match well with that obtained using the method in Ref.~\citen{Quan2018}. In addition, the  parameters retrieved using the present approach automatically satisfy the constraints imposed by reciprocity and energy conservation.

This paper is arranged as follows. In Sec.~\ref{scath}, we formulate the scattered fields for both 2D and 1D problems. Then we present our retrieval method in Sec.~\ref{retr} with Eqs.~\eqref{mod2d} and \eqref{retralphap} being the main results. Numerical examples are shown in Sec.~\ref{examp1} to validate our method. Section \ref{Conc} concludes the paper.

\section{\label{scath}Scattering from sub-wavelength Willis acoustic elements}

The type of scatterer considered here is one that couples monopole and dipole terms with little or no contribution from higher order multipoles.  In this way it is the simplest embodiment of a Willis material,  analogous to a standard lumped element in uni-dimensional acoustics.  For that reason we refer to the scatterer as a Willis scatterer or a Willis element. 
The scattered fields from a Willis element under arbitrary incidence can be written in terms of the polarizability tensor as first introduced by~\citet{Quan2018}. Explicit expressions for the pressure fields can be found by relating them to the multipole components of the scattered wave.  In this section we derive  the equations for the scattered fields using the polarizability tensor  and reveal several general properties. 

Consider sound radiation from a point source in a background medium with mass density $\rho$ and sound speed $c$. The radiated pressure can be written in terms of the Green's function as
\begin{equation}\label{scatgre}
p_s({\bf x})=\omega^2MG({\bf x})-\omega^2{\bm {D}}\cdot \nabla G({\bf x})+\cdots ,
\end{equation}
where $\omega$ is the angular frequency $(e^{-i\omega t}$ assumed).  All  high order multipole terms are dropped in the multipole expansion  except for the monopole mass $M$ and dipole moment $\bm D$. The Green's function $G({\bf x})$ takes the form
\begin{equation}\label{gf}
\begin{aligned}
G({\bf x})=\begin{cases}
\frac{1}{i2kS}e^{ik|{\bf x}|}, &{\text {1D waveguide}},\\
\frac{1}{i4}H_0^{(1)}(k|{\bf x}|), &{\text {2D free field}},\\
\frac{-1}{4\pi|{\bf x}|}e^{ik|{\bf x}|} &{\text {3D free field}},
\end{cases}
\end{aligned}
\end{equation}
where $S$ is the cross-sectional area of the 1D waveguide; $k=\omega /c$ is the wavenumber in the background medium; $H_n^{(1)}$ is the Hankel function of the first kind. The mass  and  dipole moment are proportional to the incident pressure $p_i$ and velocity ${\bm v_i}$,   
\begin{equation}\label{md}
\begin{aligned}
\begin{pmatrix}
M\\ {\bm D}
\end{pmatrix}={\bm \alpha}
\begin{pmatrix}
p_i\\ {\bm v_i}
\end{pmatrix},
\end{aligned}
\end{equation}
where  ${\bm \alpha}$ is  the polarizability tensor with components 
\begin{equation}\label{tensor}
\begin{aligned}
{\bm \alpha}=\begin{pmatrix}
\alpha^{pp} &{\bm \alpha^{pv}}^T\\
{\bm \alpha^{vp}} &{\bm \alpha^{vv}}
\end{pmatrix}.
\end{aligned}
\end{equation}
The diagonal terms in Eq.~\eqref{tensor} correspond to the pressure excited monopole and velocity excited dipole; the off-diagonal terms correspond to the cross-coupling induced monopole and dipole moment. The objective of this paper is to provide a method to determine the components of $\bm \alpha$.
We focus on the retrieval method for the 2D and 1D cases, the 3D case can be derived in a manner similar to the 2D case.

\subsection{Scattering from a Willis element  in 2D }
Consider an acoustically small asymmetric scatterer in 2D, with incident pressure and velocity fields at the scatterer location  defined as
\begin{equation}\label{2Dinci}
\begin{aligned}
\begin{pmatrix}
p_i\\ v_{xi}\\ v_{yi}
\end{pmatrix}=
\begin{pmatrix}
1 &0 &0\\ 0 &\frac{1}{\sqrt{2}\rho c} &0\\ 0 &0 &\frac{1}{\sqrt{2}\rho c}
\end{pmatrix}
\begin{pmatrix}
A_0\\ A_x\\ A_y
\end{pmatrix}.
\end{aligned}
\end{equation}
Using Eqs.~\eqref{scatgre} and \eqref{gf}, the scattered pressure field from a Willis element is
\begin{equation}\label{sctp2d}
\begin{aligned}
p_s({\bf x})&=\frac{\omega^2}{i4}\Big ( MH_0^{(1)}(kr) + k \hat{\bf x}\cdot {\bm D}H_1^{(1)}(kr)\Big )\\
&=T_0H_0^{(1)}(kr)+\big ( T_x\cos \theta + T_y \sin \theta \big )i\sqrt{2}H_1^{(1)}(kr),
\end{aligned}
\end{equation}
where
\begin{equation}\label{tdef2d}
\begin{aligned}
{\bm t}\equiv \begin{pmatrix}
T_0\\ T_x\\ T_y
\end{pmatrix}=
\frac{\omega^2}{i4}
\begin{pmatrix}
1 &0 &0\\ 0 &-\frac{ik}{\sqrt{2}} &0\\ 0 &0 &-\frac{ik}{\sqrt{2}}
\end{pmatrix}
\begin{pmatrix}
M\\ D_x\\ D_y
\end{pmatrix}.
\end{aligned}
\end{equation}
The 2D (scattering) T-matrix ${\bm T}$ is defined by ${\bm t}={\bm T}{\bm a}$  with ${\bm a}=(A_0\ A_x\ A_y)^T$,
and hence 
\begin{align}\label{TM2d}
{\bm T}&=\frac{\omega^2}{i4}\begin{pmatrix}
1 &0 &0\\ 0 &-\frac{ik}{\sqrt{2}} &0\\ 0 &0 &-\frac{ik}{\sqrt{2}}
\end{pmatrix}{\bm \alpha}
\begin{pmatrix}
1 &0 &0\\ 0 &\frac{1}{\sqrt{2}\rho c} &0\\ 0 &0 &\frac{1}{\sqrt{2}\rho c}
\end{pmatrix}
\notag \\
&=i\frac{\omega^2}{8}{\bm \alpha}^\prime,
\end{align}
where the modified polarization tensor  ${\bm \alpha}^\prime$ is 
\begin{align}\label{modalpha2d}
{\bm \alpha}^\prime &= \begin{pmatrix}
\alpha^{pp\prime} &\alpha_x^{pv\prime} &\alpha^{pv\prime}_y\\
\alpha^{vp\prime}_x &\alpha^{vv\prime}_{xx} &\alpha^{vv\prime}_{xy}\\
\alpha^{vp\prime}_y &\alpha^{vv\prime}_{yx} &\alpha^{vv\prime}_{yy}
\end{pmatrix}
\notag \\ &=
\begin{pmatrix}
-2\alpha^{pp} &-\frac{\sqrt{2}}{\rho c}\alpha_x^{pv} &-\frac{\sqrt{2}}{\rho c}\alpha^{pv}_y\\
ik\sqrt{2}\alpha^{vp}_x &\frac{ik}{\rho c}\alpha^{vv}_{xx} &\frac{ik}{\rho c}\alpha^{vv}_{xy}\\
ik\sqrt{2}\alpha^{vp}_y &\frac{ik}{\rho c}\alpha^{vv}_{yx} &\frac{ik}{\rho c}\alpha^{vv}_{yy}
\end{pmatrix}.
\end{align} 
This expression is consistent with that in Ref.~\citen{Quan2018}. Reciprocity implies that $\alpha_{xy}^{vv\prime}=\alpha_{yx}^{vv\prime}$, $\alpha_x^{pv\prime}=-\alpha_x^{vp\prime}$ and $\alpha_y^{pv\prime}=-\alpha_y^{vp\prime}$. We will see later in the numerical examples that the retrieved parameters satisfy these reciprocity requirements.

The explicit form of the scattered field for any  incident wave can be obtained using  the above equations. For example, a plane wave incident along the $+x$-direction with unit magnitude is defined by  $A_0=A_x/\sqrt{2}=1$ and $A_y=0$; the unit plane wave incident along the positive diagonal direction corresponds to $A_0=A_x=A_y=1$.

The importance of using the T-matrix formalism is that it is  related to the S-matrix,  
${\bm S}={\bm I}+2{\bm T}$, which is unitary in the absence of absorption: ${\bm S}{\bm S}^\dagger = {\bm I}$. 
This places direct limits on the polarizability tensor, specifically that the eigenvalues of 
${\bm I}+ \frac i{4} \omega^2{\bm \alpha}^\prime $ must be of unit magnitude.  In the case of  energy dissipation the eigenvalue magnitudes must be less than or equal to unity.  This result and its implications, which  generalizes the bounds obtained by \cite{Quan2018}, will be discussed at further length separately.

\subsection{Scattering from a Willis element in a 1D waveguide}
The scattering problem in a 1D  waveguide is simpler in that it only involves forward and backward scattered waves. In this case  the incident pressure and velocity fields at the scattering location are
\begin{equation}\label{pv}
\begin{aligned}
\begin{pmatrix}
p_i\\ {v_i}
\end{pmatrix}=
\begin{pmatrix}
1 &0\\ 0 &\frac{1}{\rho c}
\end{pmatrix}\begin{pmatrix}
A_1\\ A_2
\end{pmatrix}.
\end{aligned}
\end{equation}
Using Eqs.~\eqref{scatgre} and \eqref{gf}, the scattered pressure field from a Willis element in the waveguide is
\begin{align}\label{scat}
p_s(x)&=\frac{\omega c}{i2S}\Big (M-ikD \sgn x \Big ) e^{ik|x|}\notag \\
&=\big (T_0+T_1\sgn x\big ) e^{ik|x|}.
\end{align}
Note that the size of the scatterer must be much smaller than the radius of the waveguide, so we do not need to consider the narrow region nearby. The frequency range is sufficiently low so that only the fundamental waveguide mode  propagates. By analogy with the  2D case,  
\begin{equation}\label{scattm}
\begin{aligned}
{\bm t}\equiv \begin{pmatrix}
T_0\\ T_1
\end{pmatrix}
=\frac{i\omega c}{2S}
\begin{pmatrix}
-1 &0 \\ 0 &ik
\end{pmatrix}
\begin{pmatrix}
M\\  D
\end{pmatrix}.
\end{aligned}
\end{equation}
The  scattering matrix,  defined by ${\bm t} ={\bm {T}}(A_1\ A_2)^T$, is  
\begin{equation}\label{bigt}
{\bm T}=\frac{i\omega c}{2S}
\begin{pmatrix}
-1 &0\\ 0 &ik
\end{pmatrix}{\bm \alpha}
\begin{pmatrix}
1 &0\\ 0 &\frac{1}{\rho c}
\end{pmatrix}=\frac{i\omega c}{2S}{\bm {\alpha^\prime}},
\end{equation}
and the modified polarizability tensor is  
\begin{equation}\label{ptensor}
\begin{aligned}
{\bm \alpha^\prime}=\begin{pmatrix}
\alpha^{pp\prime} &\alpha^{pv\prime}\\
\alpha^{vp\prime} &\alpha^{vv\prime}
\end{pmatrix}=\begin{pmatrix}
-\alpha^{pp} &-\frac{1}{\rho c}\alpha^{pv}\\
ik\alpha^{vp} &\frac{ik}{\rho c}\alpha^{vv}
\end{pmatrix}.
\end{aligned}
\end{equation}
Reciprocity requirements impose the  constraint: $\alpha^{pv\prime}=-\alpha^{vp\prime}$. It should be pointed out that the monopole and dipole moments are more dominant in this case since only the fundamental mode  can  propagate within the frequency range of interest. Therefore, the two eigenvalues of the S-matrix ${\bm S}={\bm I}+2{\bm T}$ must be close to unity.

The explicit form of the scattered field can be obtained as in 2D. The form of the incident waves are simpler with $A_1=A_2=1$ corresponding to a unit amplitude wave  incident wave in the $+x$-direction, while  $A_1=-A_2=1$ corresponds to a wave incident along the $-x$-direction.

\section{Retrieval method for the polarizability tensor}\label{retr}

The  coupling between monopoles and dipoles in a Willis scatterer is achieved by a physically asymmetric object, for which there are rarely closed-form expressions available for the polarizability tensor ${\bm \alpha}$.  It is therefore essential to have a rigorous and efficient numerical retrieval method. The objective of this section is to provide  a simple recipe for retrieving ${\bm \alpha}$ from FEM simulations or experimental data. 

\subsection{2D free field}

In this section we will use the nine-component form of the $3\times 3$ polarizability tensor for later comparison, even though it  has only six  independent components due to reciprocity. 
We consider four plane wave excitations along the $\pm x$- and $\pm y$-directions. These are defined by taking appropriate  amplitudes in Eq.~\eqref{2Dinci}. For instance,   a plane wave   incident along the $+x$-direction with pressure and velocity at the scatterer location (the origin) equal to  $1$  and $1/\rho c$, respectively, 
corresponds to  $A_0=A_x/\sqrt{2}=1$ and $A_y=0$. Similarly,  $A_0=-A_x/\sqrt{2}=1$ and $A_y=0$ for $-x$-incidence; $A_0=A_y/\sqrt{2}=1$ and $A_x=0$ for $y$-incidence; $A_0=-A_y/\sqrt{2}=1$ and $A_x=0$ for $-y$-incidence. 
The scattered pressure for the four cases are
\begin{widetext}
\begin{equation}\label{2dcase}
\begin{aligned}
^\pm_xp_s({\bf x}) &=\frac{i\omega^2}{8}\big (\alpha^{pp\prime}\pm\sqrt{2}\alpha_x^{pv\prime}\big )H_0^{(1)}(kr)
-\frac{\sqrt{2}\omega^2}{8}\Big [ \big (\alpha_x^{vp\prime}\pm\sqrt{2}\alpha_{xx}^{vv\prime}\big )\cos \theta + \big (\alpha_y^{vp\prime}\pm\sqrt{2}\alpha_{yx}^{vv\prime}\big ) \sin \theta \Big ]H_1^{(1)}(kr),\\
^\pm_yp_s({\bf x}) &=\frac{i\omega^2}{8}\big (\alpha^{pp\prime}\pm\sqrt{2}\alpha_y^{pv\prime}\big )H_0^{(1)}(kr)
-\frac{\sqrt{2}\omega^2}{8}\Big [ \big (\alpha_x^{vp\prime}\pm\sqrt{2}\alpha_{xy}^{vv\prime}\big )\cos \theta + \big (\alpha_y^{vp\prime}\pm\sqrt{2}\alpha_{yy}^{vv\prime}\big ) \sin \theta \Big ]H_1^{(1)}(kr),
\end{aligned}
\end{equation}
\end{widetext}
where the superscript and subscript on the left side of $p_s$ denote the incident direction. 
For example, $^-_xp_s({\bf x})$ is the  scattering solution for incidence along the $-x$-direction. 

The retrieval method uses four scattered pressure measurements at distance $r=|{\bf x}|$ along $\pm x$- and $\pm y$-directions for each excitation, implying 16 data points. 
The probed pressure values  are $^{\pm}_xp_{sx}^{\pm}(r)$, $^{\pm}_xp_{sy}^{\pm}(r)$, $^{\pm}_yp_{sx}^{\pm}(r)$ and $^{\pm}_yp_{sy}^{\pm}(r)$, with the superscript and subscript on the right side of $p_s$ denoting the location $r=|{\bf x}|$, at which the pressure is measured along the specific axis. 
For instance, $^-_xp_{sy}^{+}(r)$ is the scattered pressure at distance $r$ along the $+y$-direction for incidence along the $-x$-direction. However, the quadrupole component is not negligible when $ka\sim 1$. In order to obtain more accurate results, we filter out the quadrupole based on the orthogonality of each harmonic. For example, the quadrupole components in the four scattered fields for $+x$-incidence can be filtered by setting ${_x^+R} = ({^+_xp_{sx}^+} - {^+_xp_{sy}^+} + {^+_xp_{sx}^-} - {^+_xp_{sy}^-})/4$ and redefining the pressures as ${^+_xp_{sx}^+} - {_x^+R} \rightarrow {^+_xp_{sx}^+}$, ${^+_xp_{sx}^-} - {_x^+R} \rightarrow {^+_xp_{sx}^-}$, ${^+_xp_{sy}^+} + {_x^+R} \rightarrow {^+_xp_{sy}^+}$ and ${^+_xp_{sy}^-} + {_x^+R} \rightarrow {^+_xp_{sy}^-}$. All the other probed data corresponding to different incidences should be filtered in a similar fashion, such that they only include monopole and dipole components.
Plugging the 16 filtered pressures into Eq.~\eqref{2dcase} and omitting $(r)$ for conciseness, we may invert the extracted data to get   the modified polarizability components: 
\begin{widetext}
\begin{equation}\label{mod2d}
\begin{aligned}
\alpha^{pp\prime}     &=\frac{-i}{\omega^2H_0^{(1)}(kr)}        \Big ( {^+_xp_{sx}^-} + {^+_xp_{sx}^+} + {^+_yp_{sy}^-} + {^+_yp_{sy}^+} + {^-_xp_{sx}^-} + {^-_xp_{sx}^+} + {^-_yp_{sy}^-} + {^-_yp_{sy}^+} \Big ),\\
\alpha_x^{pv\prime}   &=\frac{-i}{\sqrt{2}\omega^2H_0^{(1)}(kr)} \Big ( {^+_xp_{sx}^-} + {^+_xp_{sx}^+} + {^+_xp_{sy}^-} + {^+_xp_{sy}^+} - {^-_xp_{sx}^-} - {^-_xp_{sx}^+} - {^-_xp_{sy}^-} - {^-_xp_{sy}^+} \Big ),\\
\alpha_y^{pv\prime}   &=\frac{-i}{\sqrt{2}\omega^2H_0^{(1)}(kr)} \Big ( {^+_yp_{sy}^-} + {^+_yp_{sy}^+} + {^+_yp_{sx}^-} + {^+_yp_{sx}^+} - {^-_yp_{sy}^-} - {^-_yp_{sy}^+} - {^-_yp_{sx}^-} - {^-_yp_{sx}^+} \Big ),\\
\alpha_x^{vp\prime}   &=\frac{1}{\sqrt{2}\omega^2H_1^{(1)}(kr)}   \Big ( {^+_xp_{sx}^-} - {^+_xp_{sx}^+} + {^-_xp_{sx}^-} - {^-_xp_{sx}^+} + {^+_yp_{sx}^-} - {^+_yp_{sx}^+} + {^-_yp_{sx}^-} - {^-_yp_{sx}^+} \Big ),\\
\alpha_y^{vp\prime}   &=\frac{1}{\sqrt{2}\omega^2H_1^{(1)}(kr)}   \Big ( {^+_yp_{sy}^-} - {^+_yp_{sy}^+} + {^-_yp_{sy}^-} - {^-_yp_{sy}^+} + {^+_xp_{sy}^-} - {^+_xp_{sy}^+} + {^-_xp_{sy}^-} - {^-_xp_{sy}^+} \Big ),\\
\alpha_{xx}^{vv\prime}&=\frac{1}{\omega^2H_1^{(1)}(kr)}          \Big ( {^+_xp_{sx}^-} - {^+_xp_{sx}^+} - {^-_xp_{sx}^-} + {^-_xp_{sx}^+} \Big ),\\
\alpha_{xy}^{vv\prime}&=\frac{1}{\omega^2H_1^{(1)}(kr)}          \Big ( {^+_yp_{sx}^-} - {^+_yp_{sx}^+} - {^-_yp_{sx}^-} + {^-_yp_{sx}^+} \Big ),\\
\alpha_{yx}^{vv\prime}&=\frac{1}{\omega^2H_1^{(1)}(kr)}          \Big ( {^+_xp_{sy}^-} - {^+_xp_{sy}^+} - {^-_xp_{sy}^-} + {^-_xp_{sy}^+} \Big ),\\
\alpha_{yy}^{vv\prime}&=\frac{1}{\omega^2H_1^{(1)}(kr)}          \Big ( {^+_yp_{sy}^-} - {^+_yp_{sy}^+} - {^-_yp_{sy}^-} + {^-_yp_{sy}^+} \Big ).
\end{aligned}
\end{equation}
\end{widetext}
Note that the combinations of the pressures in Eq.~\eqref{mod2d} for each polarizability is not unique (at least 4 out of the 16 probed pressures for each polarizability), here we took more pressure data into account to give more robust results. When implementing this retrieval method in FEM simulations, one only needs to simulate the aforementioned four plane wave excitations and extract the necessary parameters for Eq.~\eqref{mod2d}.
Thus, all the components in ${\bm \alpha}^\prime$ can be determined, and the original polarizability tensor $\bm \alpha$ follows from the relations in Eq.~\eqref{modalpha2d}.

The retrieved parameters should satisfy the constraints imposed by reciprocity:  $\alpha_x^{pv\prime}=-\alpha_x^{vp\prime}$, $\alpha_y^{pv\prime}=-\alpha_y^{vp\prime}$ and $\alpha_{xy}^{vv\prime}=\alpha_{yx}^{vv\prime}$.  These relations are in fact automatically satisfied  by the retrieval method of Eq.\ \eqref{mod2d} by virtue of  reciprocity  identities for the scattered pressure.   For instance, reciprocity under the interchange of source and receiver yields
$  {^+_xp_{sx}^+} = {^-_xp_{sx}^-}$, which combined with   Eq.\ \eqref{mod2d} implies that   $\alpha_x^{pv\prime}=-\alpha_x^{vp\prime}$.   The identity $\alpha_y^{pv\prime}=-\alpha_y^{vp\prime}$ follows in the same way from   
the relation $  {^+_yp_{sy}^+} = {^-_yp_{sy}^-}$.  Finally,  $\alpha_{xy}^{vv\prime}=\alpha_{yx}^{vv\prime}$ is a consequence of  four reciprocal relations for the scattered pressure: 
${^+_yp_{sx}^-}={^+_xp_{sy}^-}$,  ${^-_yp_{sx}^+} ={^-_xp_{sy}^+}$, 
${^+_yp_{sx}^+} = {^-_xp_{sy}^-} $ and $ {^-_yp_{sx}^-} = {^+_xp_{sy}^+}$. 


\subsection{1D waveguide}
The retrieval procedure for Willis scatterers in a 1D waveguide is  simpler than the previous 2D case since  there are only four polarizability components to be found.  Consider a wave  incident along the $\pm x$-direction (axial direction) such that the incident pressure and velocity  at the scatterer location are $1$ and $\pm 1/\rho c$, respectively. Hence, taking $A_1=\pm A_2=1$, the scattered fields for these two incidences are, with obvious notation, 
\begin{equation}\label{scat1dp}
^\pm p_s(x) =\frac{i\omega c}{2S}\Big [ \alpha^{pp\prime}\pm \alpha^{pv\prime}+
\big( \alpha^{vp\prime} \pm  \alpha^{vv\prime} \big) \sgn x \Big ]e^{ik|x|}. 
\end{equation}
Following the 2D  procedure, the polarizabilities can be expressed in terms of the forward and backward scattered pressure  as
\begin{equation}\label{retralphap}
\begin{pmatrix}
\alpha^{pp\prime} \\
\alpha^{pv\prime} \\
\alpha^{vp\prime} \\
\alpha^{vv\prime}
\end{pmatrix}
= \frac{Se^{-ik|x|}}{2i\omega c }
\begin{pmatrix}
 1 &  1 &  1 &  1 
\\
 1 &  1 &  -1 &  -1 
\\
 1 &  -1 &  1 &  -1 
\\
 1 &  -1 &  -1 &  1 
\end{pmatrix}
\begin{pmatrix}
{^+p_{s}^+} \\ 
{^+p_{s}^-} \\
{^-p_{s}^+} \\
{^-p_{s}^-}
\end{pmatrix} .
\end{equation}
Hence,  in order to retrieve the polarizabilities we only need to simulate plane wave incidence from the two opposite directions and probe on both sides of the scatterer.

The retrieved parameters should satisfy the reciprocity relation $\alpha^{pv\prime}=-\alpha^{vp\prime}$.  It is clear from Eq.\ \eqref{retralphap} that this is equivalent to the identity $^+p_{s}^+ = ^-p_{s}^-$, which is guaranteed by invariance under the interchange of source and receiver.    This implies that the transmission coefficient $T$ is independent of the direction of incidence, where $T$  is defined such that  $^+p_{s}^+ = ^-p_{s}^- = (T-1)e^{ik|x|}$.   The related 
reflection coefficients $R_{\pm}$ are defined by $^+p_{s}^- =R_+e^{ik|x|}$ and $^-p_{s}^+ =R_-e^{ik|x|}$.
Using $\alpha^{pv\prime}=-\alpha^{vp\prime}$, we have 
\begin{equation}\label{tr}
\begin{aligned}
T &= 1+\frac{i\omega c}{2S}\Big (\alpha^{pp \prime}+\alpha^{vv\prime}\Big ),\\
R_{\pm} &= \frac{i\omega c}{2S}\Big (\alpha^{pp\prime} \pm \alpha^{pv\prime}-\alpha^{vv\prime}\Big ) . 
\end{aligned}
\end{equation}
This form is similar to the transmission and reflection coefficients for two-dimensional bianisotropic materials under normal incidence in electromagnetics.\cite{Yazdi2015} 
Clearly,  as an alternative to Eq.\ \eqref{retralphap} one may  write the polarizability tensor in terms of the transmission and reflection coefficients,  
\begin{equation}\label{2=1}
\begin{pmatrix}
\alpha^{pp\prime} \\
\alpha^{vv\prime} \\
\alpha^{pv\prime} 
\end{pmatrix}
= \frac{S}{2i\omega c }
\begin{pmatrix}
 2 &  1 &  1  
\\
 2 &  -1 &  -1  
\\
 0 &  2 &  -2 
\end{pmatrix}
\begin{pmatrix}
T-1 \\ 
R_+ \\
R_-
\end{pmatrix}  
\end{equation}
with $  \alpha^{vp\prime} = - \alpha^{pv\prime}$.

The asymmetric reflection for waves incident from opposite directions is a characteristic  property of Willis elements, which is not observed for similarly sub-wavelength  monopole or dipole scatterers.  
 The difference in the reflection coefficients is induced by the pressure-velocity cross-coupling term $\alpha^{vp\prime}$ $( = - \alpha^{pv\prime})$. More specifically, the pressure excited dipole and the velocity excited monopole  interfere destructively  in the forward direction, but  constructively  for the backward scattered wave. In the absence of material loss, the two reflected waves have the same magnitude but different phases. For lossy Willis scatterers, the  magnitudes are unequal because the momentum exchange processes and hence the absorption depends on the direction of incidence. This feature can potentially be used to design unidirectional perfect absorbers or unidirectional reflectionless materials.

\section{Numerical examples}\label{examp1}
 Two examples  are presented  to demonstrate how our retrieval method works in 2D free field and 1D waveguide situations. The common procedure is to simulate  plane wave excitation and probe the scattered pressure according to the algorithms presented in Sec.~\ref{retr}, then use Eq.~\eqref{mod2d} or Eq.~\eqref{retralphap} to calculate  the components of the modified polarizability tensor.

\subsection{2D free field}
\begin{figure*}[ht] 
 \includegraphics[width=0.95\textwidth]{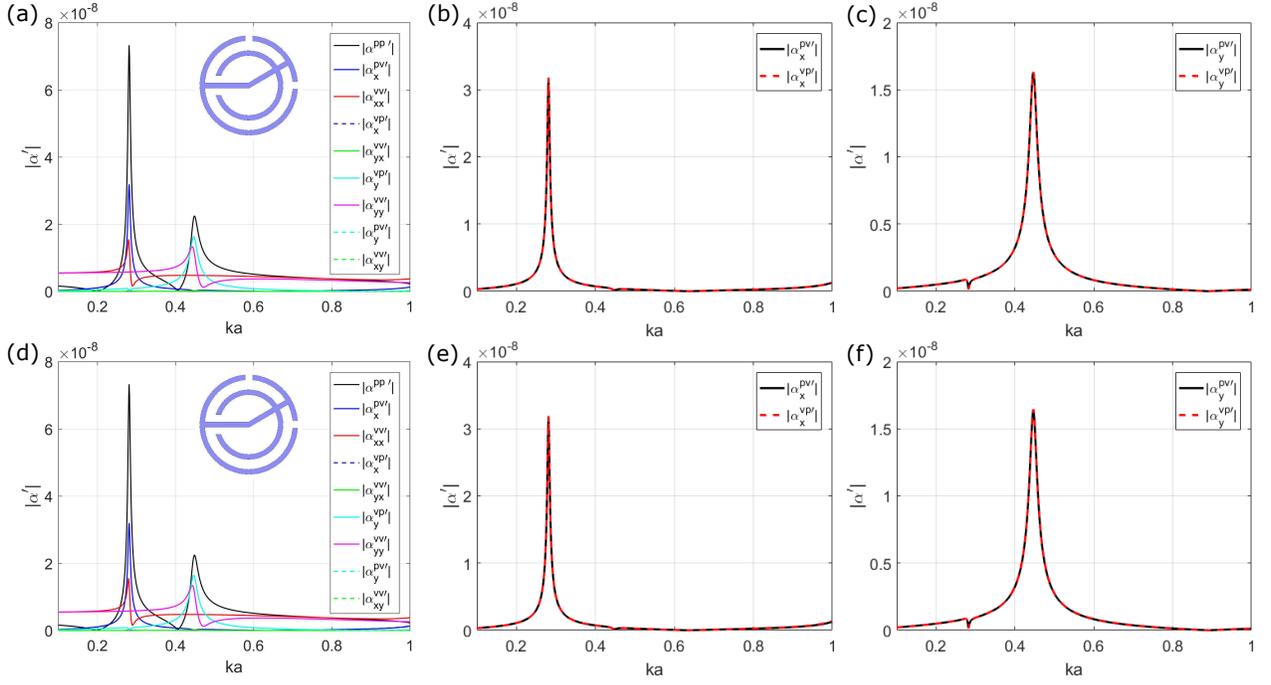} 
 \caption{Retrieved polarizability components for the 2D Willis scatterer. The nine polarizabilities are shown in (a) and (d); the cross-coupling terms are shown in (b), (c), (e) and (f). The curves in (a), (b) and (c) are obtained using the present retrieval method; the curves in (d), (e) and (f) are calculated using the method of \citet{Quan2018}.} \label{ret2dexa}
\end{figure*}

We consider a Willis element  with a radius on the order of $\lambda/10$, where $\lambda$ is the wavelength in the background medium, see Fig.~\ref{ret2dexa}(a). The wall of the scatterer is  acoustically rigid. The scatterer consists of two separate cavities with  openings in separate directions and of different volume so that the scatterer does not display any symmetry. In this way, the cross-coupling induced scattering is non-zero. Our objective here is not to maximize the cross-coupling, but rather to demonstrate that the retrieval method works when the  off-diagonal terms are on the same order of magnitude as the diagonal terms.

The background medium is air with bulk modulus $B=1.42\times 10^5$ Pa and mass density $\rho=1.225$ kg/m$^3$. The acoustically rigid and small Willis scatterer was placed at the origin of the Cartesian coordinate system. Four plane wave excitations along the $\pm x$- and $\pm y$-directions were simulated. Then four probes of the scattered pressure were taken at a fixed  distance $r$ from the origin along the $\pm x$- and $\pm y$-directions for each excitation, thus providing  the sixteen scattered pressure data needed for the parameter retrieval. 
Plugging the probed pressure into Eq.~\eqref{mod2d} yields the nine components of the modified polarizability tensor.  The full wave FEM simulations were performed using COMSOL Multiphysics. 

\begin{figure}[ht] 
\centering
 \includegraphics[width=.95\columnwidth]{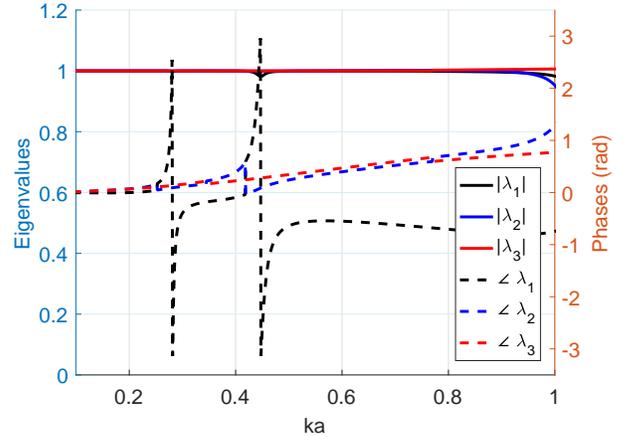}
 \caption{Eigenvalues of the S-matrix. The solid lines represent the absolute values of the eigenvalues; the dashed lines are the corresponding phases.} \label{2deig}
\end{figure}

Figure~\ref{ret2dexa} shows the frequency dependence of the polarizability components from $ka=0.1$ to $ka=1$, where $a$ is the scatterer radius. The results in panels (a), (b) and (c) are calculated using the retrieval method developed in this paper; the curves in panels (d), (e) and (f) are obtained using the method presented by \citet{Quan2018}, which requires an infinite set of data as compared with the small data set used here. It is clear that the results obtained by these two methods match to a remarkable degree. Figures \ref{ret2dexa}(a) and \ref{ret2dexa}(d) show all  nine components of the polarizability tensor. It can be seen that the $\alpha^{pp\prime}$ component responsible for the pressure excited monopole is on the same order of magnitude as the the cross-coupling terms $\alpha_x^{pv\prime}$, $\alpha_x^{vp\prime}$, $\alpha_y^{pv\prime}$ and $\alpha_y^{vp\prime}$.  
The off-diagonal terms satisfy the constraints imposed by reciprocity, i.e. $\alpha_x^{pv\prime}=-\alpha_x^{vp\prime}$, $\alpha_y^{pv\prime}=-\alpha_y^{vp\prime}$ and $\alpha_{xy}^{vv\prime}=-\alpha_{yx}^{vv\prime}$, indicating that the numerical simulation is physically consistent. The plots in this section only show the components of the modified polarization tensor ${\bm \alpha}^\prime$, one may also calculate ${\bm \alpha}$ using Eq.~\eqref{modalpha2d}. The retrieval procedure can be used to analyze more complicated scatterers including loss effects.
The T-matrix formalism has a special implication on the S-matrix, ${\bm S} = {\bm I}+2{\bm T}$, that its eigenvalue magnitudes must be less than or equal to unity due to energy conservation. In the absence of absorption, the retrieved polarizability components lead to three unitary eigenvalues for the S-matrix as shown in Fig.~\ref{2deig}. It is clear that the magnitudes are close to unity at the low frequency range, and start to decrease when $ka\sim 1$ since higher order multipoles come into play.

\subsection{1D waveguide}
\begin{figure}[ht] 
\centering
 \includegraphics[width=.95\columnwidth]{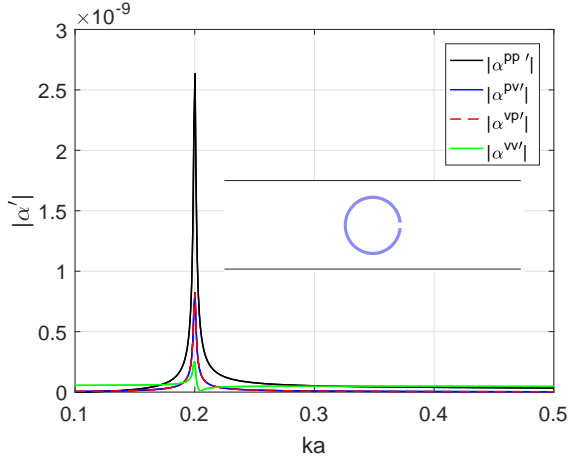}
 \caption{Retrieved polarizability components for the 1D Willis scatterer in a circular waveguide (side view; the size of the scatterer is exaggerated).} \label{1dexamp}
\end{figure}

We consider an acoustically small Willis scatterer centered in a circular rigid waveguide. The radius of the waveguide is much larger than the radius of the scatterer, and only the fundamental mode is allowed to propagate within the frequency range of interest. The scatterer has rotational symmetry about the waveguide axis, with  cross-sectional view shown in Fig.~\ref{1dexamp}.  As we can see, the Willis element is simply a spherical Helmholtz resonator which is usually considered as a monopole scatterer. However, the scattered field from such a resonator, even though deeply sub-wavelength, does depend on the direction of  incidence. 
This type of  directional scattering is more evident if the resonator is asymmetrically positioned in a waveguide, as in Fig.~\ref{1dexamp}. As we will see, the directional scattering dependence  can be attributed to the cross-coupling between the  monopole and dipole modes.

\begin{figure}[ht] 
\centering
 \includegraphics[width=.95\columnwidth]{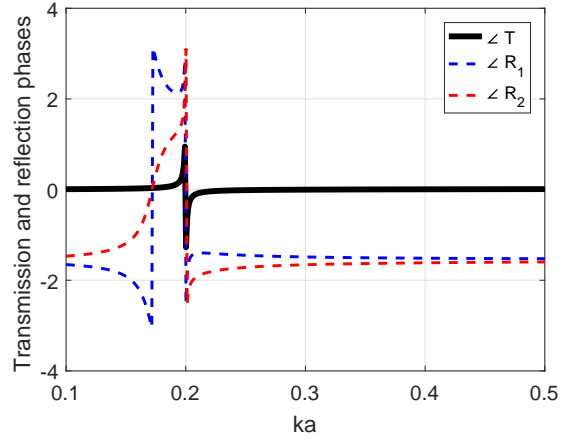}
 \caption{Phases of the transmitted and reflected waves for the two incident directions calculated using Eq.~\eqref{tr}.} \label{1dexamp2}
\end{figure}

Full wave FEM simulations (2D axisymmetric) were performed  to retrieve the polarizability tensor. The background material properties are the same as  in the previous example, and the scatterer and the waveguide are both acoustically rigid. In the 1D case, we only need to simulate plane wave incidence from each side of the scatterer. Then we probe the scattered pressure on both sides of the scatterer. Plugging the measured pressures into Eq.~\eqref{retralphap} yields the four modified polarizability components as shown in Fig.~\ref{1dexamp}. Here the off-diagonal terms also satisfy the reciprocity constraint $\alpha^{pv\prime}=-\alpha^{vp\prime}$. It is obvious that the cross-coupling terms are on the same order of magnitude compared with the pressure excited monopole and velocity excited dipole. Due to the directional dependence of the cross-coupling terms, the forward scattered fields are identical for the two incidences but the backward scattered fields are different. In the absence of material loss, the cross-coupling only contribute to different phase changes in the reflected waves. As shown in Fig.~\ref{1dexamp2}, the transmitted phases are the same for two incident directions but the reflected phases are evidently different.

\begin{figure}[ht] 
\centering
 \includegraphics[width=.95\columnwidth]{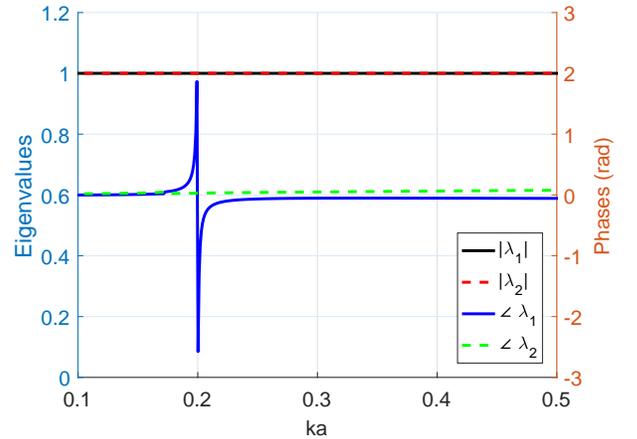}
 \caption{Eigenvalues of the S-matrix. The black and dashed red lines correspond to the absolute value of the eigenvalues; the blue and dashed green lines represent the phases of the eigenvalues.} \label{1dexamp3}
\end{figure}
As  mentioned earlier, the monopole and dipole moment are extremely dominant in the waveguide at low frequencies.  Hence the retrieved polarizabilities should lead to two unitary eigenvalues for the S-matrix ${\bm S} = {\bm I}+2{\bm T}$, which is verified in  Fig.~\ref{1dexamp3}. This  indicate that most of the energy is contained in the monopole and dipole scattering, and that  higher order multipoles are negligible in such a system.   Finally, it is worth  mention that the method developed here also works for lossy scatterers which displays more interesting phenomena such as asymmetric absorption.~\cite{Yazdi2015}

\section{Conclusion}\label{Conc}
We have presented a simple retrieval method for extracting the polarizability tensors for 2D and 1D Willis elements using a finite set of scattering amplitudes. Two examples have been presented to show the implementation procedure.  The retrieval method is based on the assumption that only monopole and dipole moment contribute to the far-field, this reduces the T-matrix to a closed $(d+1)\times(d+1)$ matrix where $d$ is the dimension. It can be seen from the 2D example that the retrieved parameters agree well with those obtained using full  monopole and dipole integration. In addition, the eigenvalues of the S-matrix in both examples are close to unity satisfying the energy conservation requirement. Therefore, our method can be used to evaluate acoustically small Willis scatterers effectively.  Although the retrieval method for 3D scattering is not presented, the derivation is straightforward following the procedure for the 2D case. The retrieval method  presented in this paper is also suitable for experimental realizations. In conclusion, the retrieval method developed in this paper can be used to design and optimize Willis inclusions for advanced wave-steering and sound absorption applications.

\section*{Acknowledgments}
This work was supported by the Office of Naval Research through MURI Grant No.\ N00014-13-1-0631.

\section*{References}


\end{document}